\documentclass[a4paper,10pt,twoside]{cpc-hepnp}

\usepackage{multicol}
\usepackage{graphicx}
\usepackage{booktabs}
\usepackage{amssymb,bm,mathrsfs,bbm,amscd}
\usepackage[tbtags]{amsmath}
\usepackage{lastpage}

\begin{document}

\fancyhead[co]{\footnotesize ZHAO Yu-Ning et al: Analytic modeling of instabilities driven by higher-order modes in HLS II RF system with a higher-harmonic cavity}

\footnotetext[0]{Submitted to 'Chinese Physics C'}

\title{Analytic modeling of instabilities driven by higher-order modes in HLS II RF system with a higher-harmonic cavity\thanks{Supported by National Natural Science
Foundation of China (10979045,11175180,11175182) }}

\author{%
\quad ZHAO Yu-Ning
\quad LI Wei-Min
\quad WU Cong-Feng$^{1)}$\email{cfwu@ustc.edu.cn}
\quad WANG Lin%
}
\maketitle

\address{%
National Synchrotron Radiation Lab., University of Science and Technology of China, Hefei 230029, China\\
}

\begin{abstract}
The utility of a passive fourth-harmonic cavity plays key role in suppressing longitudinal beam instabilities in the electron storage ring and lengthens the bunch by a factor of 2.6 for the phase II project of Hefei Light Source(HLS II). Meanwhile, instabilities driven by higher-order modes(HOM) may limit the performance of the higher-harmonic cavity. In this paper, the parasitic coupled-bunch instability which is driven by narrow band parasitic modes and the microwave instability which is driven by broadband HOM are both modeled analytically. The analytic modeling results are in good agreement with that of our previous simulation study and indicate that the passive fourth-harmonic cavity suppresses parasitic coupled-bunch instabilities and the microwave instability. The modeling suggests that a fourth-harmonic cavity may be successfully used at HLS II.
\end{abstract}



\begin{multicols}{2}

\section{Introduction}

In the phase II project of Hefei Light Source, one passive fourth-harmonic cavity will be added to the storage ring. It can help increase beam lifetime dominated by large-angle intrabeam(Touschek) scattering without diminishing the transverse beam brightness.

We have already studied some side effects induced by the application of this 816-MHz passive fourth-harmonic cavity such as Robinson instabilities and some other unstable behavior. Analytic Modeling and simulations have been used in our research. The two methods complement each other from different perspectives in which the analytic model describes the mechanism of instability, while the energy spread in simulations shows the severity. Both indicated that tuning in the harmonic cavity strongly suppresses longitudinal beam instabilities and extends the Touschek lifetime.

In this paper, we consider the effect of higher-order modes upon the longitudinal beam instability by  analytic modelings. Parasitic coupled-bunch instabilities and the microwave instability are included in our research. The parasitic coupled-bunch instability is driven by narrow band parasitic modes and the microwave instability is driven by broadband HOM. In the modeling of the parasitic coupled-bunch instability, we run the algorithm with a typical parasitic mode where the HOM resonant angular frequency equals 1000MHz. For the microwave instability, broadband impedances with the range from $0.125\Omega$ to $32\Omega$ are modeled and analytic predictions of microwave instability are given by the Boussard criterion. Compared to our previous studies, analytic modeling  results have a good agreement with simulations. Both parasitic coupled-bunch instabilities and the microwave instability are suppressed by the use of the higher-harmonic cavity. When the higher-harmonic cavity is tuned for optimal bunch lengthening, the parasitic coupled-bunch instability is strongly suppressed. The microwave instability may occur and become more severe with the increase of the broadband impedance and we figure out the broadband impedance range in which the microwave instability is not expected.

\section{Parasitic coupled-bunch instability modeling}

The parasitic coupled-bunch instability and the microwave instability can be driven by a higher-order mode. When $Q_{3}>\frac{\omega_{3}T_{0}}{8\pi}$ the HOM impedance may be excited by a single rotation sideband to cause the parasitic coupled-bunch instability\cite{lab1}, where $Q_{3}$ is the HOM quality factor, $\omega_{3}$ is the HOM resonant angular frequency and $T_{0}$ is the recirculation time. The worst-case happens when the resonant frequency of the damped HOM is an integral multiple of the revolution frequency. We consider an HOM resonant angular frequency $\omega_{3}=1000MHz$ that equals 220 times of the revolution frequency, with an HOM resonant impedance $R_{3}=10k\Omega$. For this case, the coupled-bunch growth rate is given by
\begin{equation}
\label{eq2}
|\Delta\Omega_{CB}|=\frac{eI\alpha\omega_{3}R_{3}F_{3}^{2}}{2ET_{0}\omega_{R}}.
\end{equation}
\begin{center}
\tabcaption{ \label{tab1}  The machine parameters for HLS II.}
\footnotesize
\begin{tabular*}{80mm}{c@{\extracolsep{\fill}}ccc}
\toprule Parameter &  Value \\
\hline
beam energy/Gev\hphantom{00} & 0.8 \\
beam revolution frequency/MHz\hphantom{00} & 4.533 \\
number of bunches\hphantom{00} & 45\\
synchronous voltage/kV\hphantom{00} & 16.73\\
natural relative energy spread\hphantom{00} & 0.00047\\
fundamental rf angular frequency/MHz\hphantom{00} & 204 \\
fundamental cavity shunt impedance/$M\Omega$\hphantom{00} & 3.3\\
fundamental quality factor\hphantom{00} & 28000\\
fundamental cavity coupling coefficient\hphantom{00} & 2\\
harmonic-cavity harmonic number\hphantom{00} & 4\\
harmonic-cavity shunt impedance/$M\Omega$\hphantom{00} & 2.5\\
harmonic-cavity quality factor\hphantom{00} & 18000\\
harmonic-cavity coupling coefficient\hphantom{00} & 0\\
momentum compaction\hphantom{00} & 0.02\\
fundamental rf peak voltage/kV\hphantom{00} & 250\\
harmonic frequency/MHz\hphantom{00} & 816\\
radiation-damping time constant/ms\hphantom{00} & 10\\
HOM quality factor\hphantom{00} & 3000\\
\bottomrule
\end{tabular*}
\end{center}
Here, $I$ is the ring current, $\alpha$ is the momentum compaction, $E$ is the ring energy, $\omega_{R}$ is the calculated frequency of collective dipole oscillations and $F_{3}$ is the bunch form factor at frequency $\omega_{3}$, given by
\begin{equation}
\label{eq2}
F_{3}=\exp(-\omega_{3}^{2}\sigma_{t}^{2}/2).
\end{equation}
$\sigma_{t}$ is the bunch length. If $|\Delta\Omega_{CB}|-\tau_{L}^{-1}>|\Delta\Omega|_{thresh}$, we consider that Landau damping is not sufficient to prevent the parasitic coupled-bunch instability. $\tau_{L}^{-1}$ is the radiation damping rate and $|\Delta\Omega|_{thresh}$ is the dipole Laudau damping rate\cite{lab2,lab3}. The parameters we use are shown in Table~\ref{tab1}. The modeling results of the parasitic coupled-bunch instability for a given ring current and harmonic cavity tuning angle are shown in Fig.~\ref{fig1}. For a tuning angle with the range from $-83.40^{\circ}$ to$-89.80^{\circ}$ the parasitic coupled-bunch instability is predicted to occur before optimal bunch lengthening is obtained. The results are in good agreement with ones of our previous simulations in which an optimal bunch lengthening curve described by parameters of tuning angle and current is obtained and along this curve, the natural relative energy spread is not predicted\cite{lab4}.  Our simulation also shows that the energy spread exceeds its natural value by (10-30)\% before optimal bunch lengthening  is obtained. Therefore, the parasitic coupled-bunch instability is strongly suppressed when the higher-harmonic cavity is tuned for optimal bunch lengthening and we conservatively estimate that at least 10\% is suppressed in the value of the relative energy spread.
\begin{center}
\includegraphics[width=8cm]{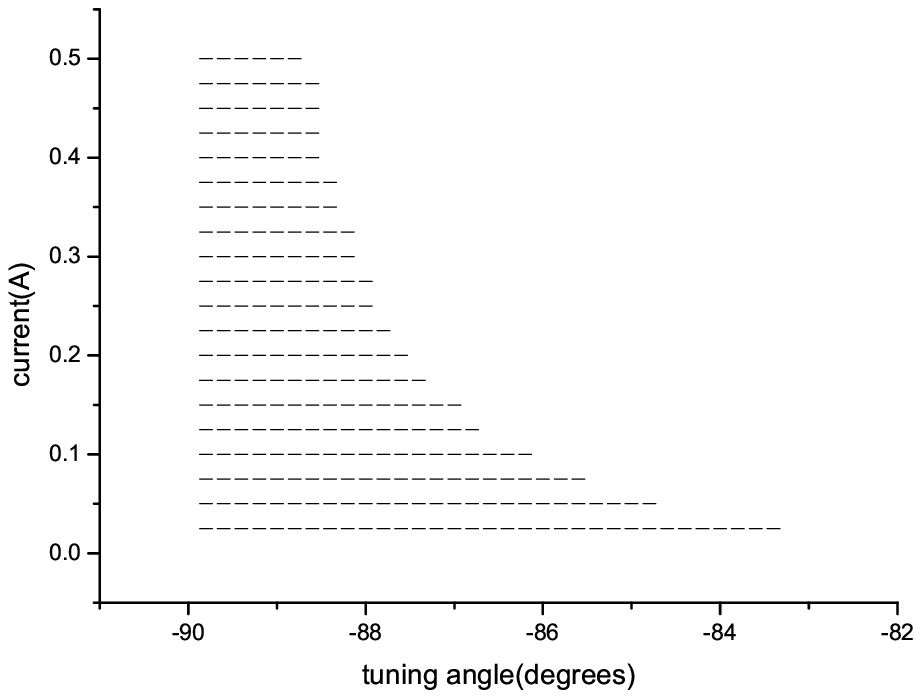}
\figcaption{\label{fig1}  Instability modeling for passive operation at HLS, in which the worst-case parasitic coupled-bunch instability is considered. Horizontal line: the parasitic coupled-bunch instability.}
\end{center}

\section{Microwave instability modeling}

Microwave instability can be considered when $Q_{3}<\frac{\omega_{3}T_{0}}{8\pi}$. For a HOM characterized by $\omega_{3}$, $R_{3}$ and $Q_{3}$, the Boussard criterion\cite{lab5} gives an approximate threshold ring current of microwave instability
\begin{equation}
\label{eq2}
I_{threshold}=\frac{\omega_{g}}{\sqrt{2\pi}}\frac{E}{e}T_{0}|\alpha|(\frac{\sigma_{E}}{E})^{2}\times\frac
{[1+Q_{3}^{2}(\omega_{3}\sigma_{t}-\frac{1}{\omega_{3}\sigma_{t}})^{2}]^{\frac{1}{2}}}{R_{3}}.
\end{equation}
$\omega_{g}$ is the fundamental rf angular frequency and $\frac{\sigma_{E}}{E}$ is the natural relative energy spread.

In our modeling, we compute the bunch length without considering any potential-well distortion from the broadband HOM. So the bunch length $\sigma_{t}$ obeys $\sigma_{t}^{2}=<\tau^{2}>-<\tau>^{2}\approx<\tau^{2}>$, where
\begin{equation}
\label{eq2}
<\tau^{n}>=\frac{\int\tau^{n}\exp[-U(\tau)/2U_{0}]d\tau}{\int\exp[-U(\tau)/2U_{0}]d\tau}.
\end{equation}
$U(\tau)$ is the Taylor expansion of the synchrotron potential and $U_{0}=\alpha^{2}(\sigma_{E}/E)^{2}/2$.
\begin{center}
\includegraphics[width=8cm]{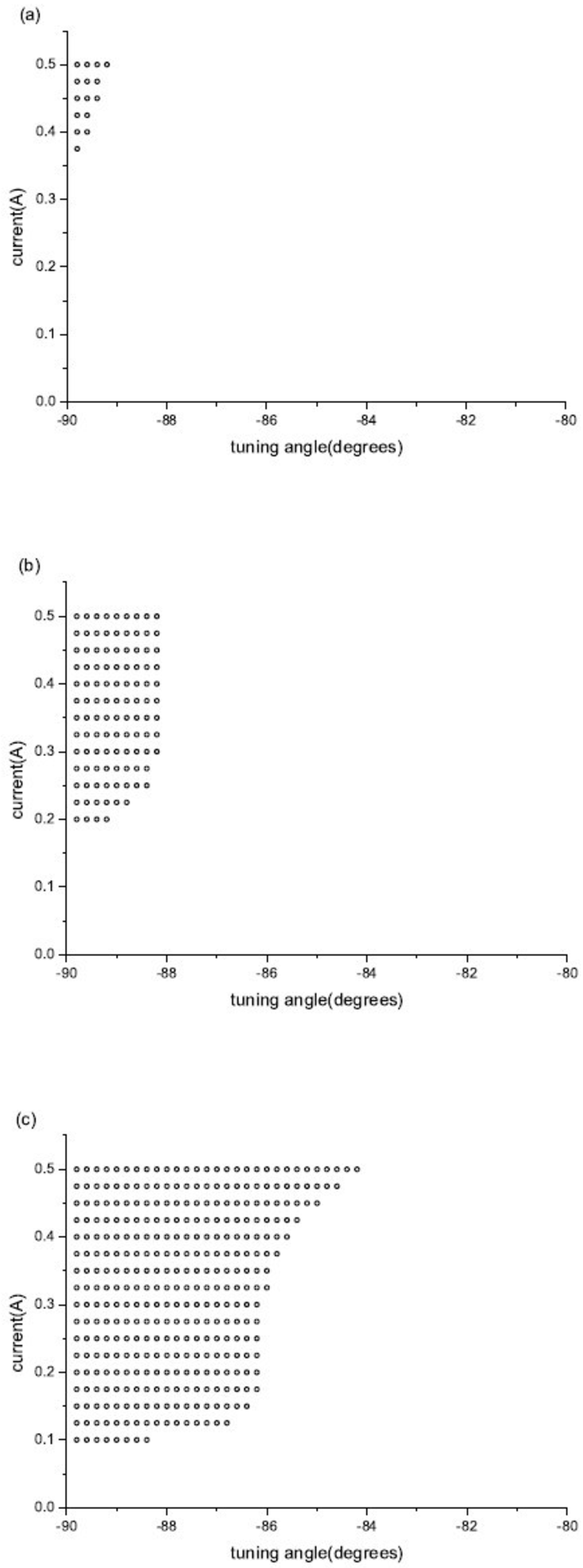}
\figcaption{\label{fig2}    Instability modeling for passive operation at HLS II with an HOM representing a broadband impedance $|Z_{p}/p|$. (a) $|Z_{p}/p|=4\Omega$, $\circ$ : microwave instability. (b) $|Z_{p}/p|=8\Omega$, $\circ$ : microwave instability. (c) $|Z_{p}/p|=16\Omega$, $\circ$ : microwave instability.    }
\end{center}

At HLS II, the microwave instability may driven by the reduced longitudinal broadband impedance $|Z_{p}/p|$ of the vacuum chamber. And only in the long-bunch regime($\omega_{3}\sigma_{t}>1$), microwave instability can be  suppressed by lengthening the bunch with a higher-harmonic cavity, because the additional bunch lengthening increases the microwave instability threshold. Therefore, we conservatively let the broadband impedance be modeled by a HOM with the smallest possible mode frequency $\omega_{3}=\omega_{c}=c/b=3183MHz$. Here, $\omega_{c}$ is the cutoff angular frequency, $c$ is the speed of light and $b=30mm$ is the average radius of a round vacuum chamber that approximates the HLS II vacuum chamber's 20-mm half height and 40-mm half width. The quality factor $Q_{3}=1$, while $|Z_{p}/p|=1\Omega$ the HOM resonant impedance is $R_{3}=351.1\Omega$. Broadband impedances with $0.125\Omega\leq|Z_{p}/p|\leq32\Omega$ are modeled and each modeling considers an impedance twice as large as the previous one. The modeling results of the microwave instability for a given ring current and harmonic cavity tuning angle are shown in Fig.~\ref{fig2}.

In our analytic modeling, we discover no microwave instability for the broadband impedance $|Z_{p}/p|\leq2\Omega$ with currents up to 500mA. Modeling results shown in Fig.~\ref{fig2} predict that if the broadband impedance $|Z_{p}/p|>2\Omega$,the microwave instability may occur and become more severe with the increasing broadband impedance. When the broadband impedance reaches up to $32\Omega$, nearly the whole range will be dotted with the microwave instability. Therefore, we would prevent the microwave instability throughout our operating current 100-500mA by decreasing the broadband impedance. We also find that the microwave instability happens first at high ring currents and it has been confirmed by simulations in our previous studies.

\section{Conclusion and discussion}

We have studied the parasitic coupled-bunch instability and the microwave instability with a higher-harmonic cavity using analytic modeling and their occurrence is approximately described by the analytic predictions. When the higher-harmonic cavity is tuned for optimal bunch lengthening, the parasitic coupled-bunch instability can be strongly suppressed. Compared with simulation results, at least 10\% is suppressed in the value of the relative energy spread.  Tuning in the higher harmonic cavity can lengthen the bunch and this additional bunch lengthening increases the microwave instability threshold, so that it suppresses the microwave instability which does not occur until the ring operates with some high currents. In our modeling, if the broadband impedance $|Z_{p}/p|\leq2\Omega$, the bunches with currents up to 500mA will not suffer from the microwave instability.

\end{multicols}

\vspace{-1mm}
\centerline{\rule{80mm}{0.1pt}}
\vspace{2mm}

\begin{multicols}{2}

\end{multicols}

\clearpage

\end{document}